\documentclass[12pt,english]{article}
\usepackage{babel}
\usepackage{amsfonts,epsfig}
\title{Infrared non-perturbative QCD running coupling    
from Bogolubov approach}
\author{Boris A. Arbuzov\\
{\it Skobeltsyn Institute of Nuclear Physics of 
MSU,}\\ {\it 119992 Moscow, RF}\\ 
E-mail: arbuzov@theory.sinp.msu.ru
}

\date{}
\newcommand{\be}{\begin{equation}}
\newcommand{\ee}{\end{equation}}
\newcommand{\beq}{\begin{eqnarray}}
\newcommand{\eeq}{\end{eqnarray}}       
\newcommand{\nn}{\nonumber}
\newcommand{\bi}{\bibitem}
\begin{document}
\maketitle
\begin{quote}
 
We apply Bogolubov approach to QCD 
to demonstrate a  spontaneous generation of three-gluon gauge invariant 
effective interaction which contributes significantly at the infrared 
region. The contribution of this interaction to $\alpha_s(p^2)$ leads to 
description of its behaviour in correspondence to phenomenology and lattice 
calculations without introduction of any additional parameter. 
\end{quote}


In works~\cite{Arb05, AVZ}  with  the use of  method~\cite{Arb04} inspired 
by 
N.N. Bogolubov approach~\cite{Bog2, Bog} a spontaneous generation 
of the effective interaction of  
the Nambu -- Jona-Lasinio type is demonstrated. The interaction 
contains no additional parameters but the QCD ones. Results of calculation of 
hadron low-energy parameters: $m_\pi,\, f_\pi,\, m_\sigma,\, \Gamma_\sigma,\, 
<\bar q q>$ are quite consistent.  
In the present letter we 
apply the approach for calculation of infrared behaviour of the 
QCD running coupling constant.

We start with QCD Lagrangian with three light quarks 
($u$, $d$ and $s$) with number of colours $N = 3$
\beq
& &L\,=\,\sum_{k=1}^3\biggl(\frac{\imath}{2}
\Bigl(\bar\psi_k\gamma_
\mu\partial_\mu\psi_k\,-\partial_\mu\bar\psi_k
\gamma_\mu\psi
_k\,\biggr)-\,m_k\bar\psi_k\psi_k\,+
\,g_s\bar\psi_k\gamma_\mu t^a A_\mu^a\psi_k
\biggr)\,-
\,\frac{1}{4}\,\biggl( F_{\mu\nu}^aF_{\mu\nu}^a\biggr);
\label{initial}\nn\\
& &F_{\mu\nu}^a\,=\,
\partial_\mu A_\nu^a - \partial_\nu A_\mu^a\,+g\,f_{abc}A_\mu^b A_\nu^c\,.
\eeq
where we use the standard QCD notations. In what follows we shall 
consider $m_k$ to be small enough and set them to zero. 
In accordance to the Bogolubov approach, application of 
which to such problems being described in details in work
~\cite{Arb04}, we look for a non-trivial solution of a 
compensation equation, which is formulated on the basis 
of the Bogolubov procedure {\bf add -- subtract}. Namely 
let us write down the initial expression~(\ref{initial}) 
in the following form
\beq
& &L\,=\,L_0\,+\,L_{int}\,;\nn\\
& &L_0\,=\,\frac{\imath}{2}
\Bigl(\bar\psi\gamma_
\mu\partial_\mu\psi-\partial_\mu\bar\psi\gamma_\mu\psi
\biggr)\,-\,\frac{1}{4}\,F_{0\,\mu\nu}^aF_{0\,\mu\nu}^a\,+ 
\,\frac{G}{3!}\cdot\,f_{abc}\,F_{\mu\nu}^a\,F_{\nu\rho}^b\,F_{\rho\mu}^c\,;
\label{L0}\\
& &L_{int}\,=\,g_s\,\bar\psi\gamma_\mu 
t^a A_\mu^a\psi\,-\,\frac{1}{4}\,\biggl( F_{\mu\nu}^aF_
{\mu\nu}^a - 
F_{0\,\mu\nu}^aF_{0\,\mu\nu}^a\biggr)\,-\,\frac{G}{3!}\cdot\,f_{abc}\,
F_{\mu\nu}^a\,F_{\nu\rho}^b\,F_{\rho\mu}^c\,.\label{Lint}
\eeq

Here $\psi$ is the light quarks triplet, colour summation is 
performed inside of each fermion 
bilinear combination, $F_{0\,\mu\nu} = 
\partial_\mu A_\nu - \partial_\nu A_\mu$, and notation  
 $\frac{G}{3!}\cdot \,f_{abc}\,
F_{\mu\nu}^a\,F_{\nu\rho}^b\,F_{\rho\mu}^c$ means corresponding 
non-local vertex in the momentum space
\beq
& &(2\pi)^4\,G\,\,f_{abc}\,(g_{\mu\nu} (q_\rho pk - p_\rho qk)+ g_{\nu\rho} 
(k_\mu pq - q_\mu pk)+g_{\rho\mu} (p_\nu qk - k_\nu pq)+\nn\\
& &q_\mu k_\nu p_\rho - k_\mu p_\nu q_\rho)\,F(p,q,k)\,
\delta(p+q+k)\,+...;\label{vertex}
\eeq
where $F(p,q,k)$ is a form-factor and 
$p,\mu;\;q,\nu;\;k,\rho$ are respectfully incoming momenta and Lorentz indices 
of gluons and we mean also that there are present four-gluon, five-gluon and 
six-gluon vertices according to expression for $F_{\mu\nu}^a$ (\ref{initial}).
 
Let us consider  expression~
(\ref{L0}) as the new {\bf free} Lagrangian $L_0$, 
whereas expression~(\ref{Lint}) as the new 
{\bf interaction} Lagrangian $L_{int}$. Then 
compensation conditions (see again~\cite{Arb04}) will 
consist in demand of full connected tree-gluon vertices, 
following from Lagrangian $L_0$, to be zero. This demand 
gives a non-linear equation for form-factor 
$F$.

These equations according to terminology of works
~\cite{Bog2, Bog} are called {\bf compensation equations}. 
In a study of these equations it is always evident the 
existence of a perturbative trivial solution (in our case 
$G = 0$), but, in general, a non-perturbative 
non-trivial solution may also exist. Just the quest of 
a non-trivial solution inspires the main interest in such 
problems. One can not succeed in finding an exact 
non-trivial solution in a realistic theory, therefore 
the goal of a study is a quest of an adequate 
approach, the first non-perturbative approximation of 
which describes the main features of the problem. 
Improvement of a precision of results is to be achieved 
by corrections to the initial first approximation.

Thus our task is to formulate the first approximation. 
Here the experience acquired in the course of performing 
works~\cite{Arb05, AVZ, Arb04} could be helpful. Now in view of 
obtaining the first approximation we would make the following 
assumptions.\\
1) In compensation equation we restrict ourselves by 
terms with loop numbers 0, 1.\\
2) In expressions thus obtained we perform a procedure 
of linearizing, which leads to linear integral 
equations. It means that in loop terms only one vertex 
contains the form-factor, being defined above, while 
other vertices are considered to be point-like. In
diagram form equation for form-factor $F$ is presented 
in Fig.1. Here four-leg vertex correspond to interaction of four 
gluons due to our effective three-field interaction. In our approximation we 
take here point-like vertex with interaction constant proportional 
to $g\,G$.\\
3) We integrate by angular variables of the 4-dimensional Euclidean 
space. The necessary rules are presented in papers~\cite{Arb05},\cite{Arb04}.

At first let us present the expression for four-gluon vertex
\beq
& &V(p,m,\lambda;\,q,n,\sigma;\,k,r,\tau,\,l,s,\pi) = g G \biggl(f^{amn} 
f^{ars}\Bigl(U(k,l;\sigma,\tau,\pi,\lambda)-U(k,l;\lambda,\tau,\pi,\sigma)-\nn\\& &-U(l,k;\sigma,\pi,\tau,\lambda)+
U(l,k;\lambda,\pi,\tau,\sigma)+U(p,q;\pi,\lambda,\sigma,\tau)-U(p,q;\tau,\lambda,\sigma,\pi)-\nn\\
& &-U(q,p;\pi,\sigma,\lambda,\tau)
+U(q,p;\tau,\sigma,\lambda,\pi)\Bigr)-f^{arn}\,
f^{ams}\Bigl(U(p,l;\sigma,\lambda,\pi,\tau)-\nn\\
& &-U(l,p;\sigma,\pi,\lambda,\tau)
-U(p,l;\tau,\lambda,\pi,\sigma)+
U(l,p;\tau,\pi,\lambda,\sigma)+U(k,q;\pi,\tau,\sigma,\lambda)-\nn\\
& &-U(q,k;\pi,\sigma,\tau,\lambda)
-U(k,q;\lambda,\tau,\sigma,\pi)
+U(q,k;\lambda,\sigma,\tau,\pi)\Bigr)+\nn\\
& &+f^{asn}\,
f^{amr}\Bigl(U(k,p;\sigma,\lambda,\tau,\pi)-U(p,k;\sigma,\tau,\lambda,\pi)
+U(p,k;\pi,\tau,\lambda,\sigma)-\nn\\
& &-U(k,p;\pi,\lambda,\tau,\sigma)-U(l,q;\lambda,\pi,\sigma,\tau)
+U(l,q;\tau,\pi,\sigma,\lambda)
-U(q,l;\tau,\sigma,\pi,\lambda)+\nn\\
& &+U(q,l;\lambda,\sigma,\pi,\tau)\Bigr)\biggr)\,;\label{four}\\
& &U(k,l;\sigma,\tau,\pi,\lambda)=k_\sigma\,l_\tau\,g_{\pi\lambda}-k_\sigma\,l_\lambda\,g_{\pi\tau}+k_\pi\,l_\lambda\,g_{\sigma\tau}-(kl)g_{\sigma\tau}g_{\pi\lambda}\,.\nn                          
\eeq
Here a triad $p,\,m,\,\lambda$ {\it etc} means correspondingly momentum, colour 
index, Lorentz index of the gluon.

Let us formulate compensation equations in this 
approximation. 
For {\bf free} Lagrangian $L_0$ full connected 
three-gluon vertices are to vanish. One can succeed in 
obtaining analytic solutions for the following set 
of momentum variables (see Fig. 1): left-hand legs 
have momenta  $p$ and $-p$, and a right-hand leg 
has zero momenta.
However in our approximation we need form-factor $F$ also 
for non-zero values of this momentum. We look for a solution 
with the following simple dependence on all three variables
\be
F(p_1,\,p_2,\,p_3)\,=\,F(\frac{p_1^2\,+\,p_2^2\,+\,p_3^2}{2})\,;\label{123}
\ee
Really, expression~(\ref{123}) is symmetric and it return to $F(x)$  
for $p_3=0,\,p_1^2\,=\,p_2^2\,=\,x$. We consider the representation~(\ref{123}) 
to be the first approximation and we plan to take into account the 
corresponding correction in the forthcoming study.

Now following the rules being stated above we 
obtain the following equation for form-factor $F(x)$ 
\beq
& &F(x)\,=\,-\,\frac{G^2\,N}{64\,\pi^2}\Biggl(\int_0^Y\,F(y)\,y dy\,-\,
\frac{1}{12\,x^2}\,\int_0^x\,F(y)\,y^3 dy\,
+\,\frac{1}{6\,x}\,\int_0^x\,F(y)\,
y^2 dy\,+\nn\\
& &+\,\frac{x}{6}\,\int_x^Y\,F(y)\,dy\,-\,\frac{x^2}{12}\,
\int_x^Y\,\frac{F(y)}{y}\,dy \Biggr)\,+\,\frac{G\,g\,N}{16\,\pi^2}\,
\int_0^Y\,F(y)\, dy\,+\label{eqF}\\
& &+\,\frac{G\,g\,N}{24\,\pi^2}\,\Biggl(\int_{3 x/16}^{x/4}\,\frac{(3 x- 16 y)^2 (3 x- 8 y)}{x^2 (x-8 y)}F(y)\, 
dy\,+\,\int_{x/4}^Y\,\frac{(5 x- 24 y)}{(x-8 y)}F(y)\, dy\Biggr)\,.\nn
\eeq
Her $x = p^2$ and $y = q^2$, where $q$ is an integration momentum. 
We introduce here  
an effective cut-off $Y$, which limited an infrared region where 
our non-perturbative effects act
and consider the equation at interval $[0,\, Y]$ under condition 
\be
F(Y)\,=\,0\,. \label{Y0}
\ee
We shall solve equation~(\ref{eqF}) by iterations. That is we 
expand the last line of~(\ref{eqF}) in powers of $x$ and 
take at first only constant term. Thus we have
\beq
& &F_0(x)\,=\,-\,\frac{G^2\,N}{64\,\pi^2}\Biggl(\int_0^Y\,F_0(y)\,y dy\,-\,
\frac{1}{12\,x^2}\,\int_0^x\,F_0(y)\,y^3 dy\,
+\,\frac{1}{6\,x}\,\int_0^x\,F_0(y)\,
y^2 dy\,+\nn\\
& &+\,\frac{x}{6}\,\int_x^Y\,F_0(y)\,dy\,-\,\frac{x^2}{12}\,
\int_x^Y\,\frac{F_0(y)}{y}\,dy \Biggr)\,+\,\frac{3\,G\,g\,N}{16\,\pi^2}\,
\int_0^Y\,F_0(y)\, dy\,.\label{eqF0}
\eeq
Expression~(\ref{eqF0}) provide an equation of the type which were 
studied in papers~\cite{Arb05, AVZ, Arb04}, 
where the way of obtaining 
solutions of equations analogous to (\ref{eqF0}) are described. Following 
this way we have the unique solution of equation (\ref{eqF0})
\beq
& &F_0(z) = \frac{1}{2}\Biggl(-\,\frac{G^2\,N}{64\,\pi^2}\,\int_0^Y F(y)\,y 
dy + \frac{3\,G\,g\,N}{16\,\pi^2} \int_0^Y F(y)\, dy 
\Biggr)\label{solution}\\
& & G_{15}^{31}\Bigl( z\,|^0_{1,\,1/2,\,0,\,-1/2,\,-1}\Bigr)=
\frac{1}{2\,z}-G_{04}^{30}\Bigl( z\,|1,\,1/2,\,-1,\,-1/2\Bigr)\,;\quad 
z\,=\,\frac{G^2\,N\,x^2}{1024\,\pi^2}\,.\nn
\eeq

Constants $C_1,\,C_2$ are defined by the following boundary conditions
\beq
& &\Bigl[2\,z^2 \frac{d^3\,F_0(z)}{dz^3}\,+9\,z\,\frac{d^2\,F_0(z)}{dz^2}\,+\,
\frac{d\,F_0(z)}{dz}\Bigr]_{z\,=\,z_0} = 0\,;\nn\\
& &\Bigl[2\,z^2\,\frac{d^2\, F_0(z)}{dz^2}\,+5\,z\,\frac{d\, F_0(z)}{dz}\,+\,
F_0(z) \Bigr]_{z\,=\,z_0} = 1\,;
\quad z_0\,=\,\frac{G^2\,N\,Y^2}{1024\,\pi^2}\,.
\label{bc}
\eeq

Here as well as in the previous papers we obtain the solution in terms of 
Meijer G-functions~\cite{be}. Conditions~(\ref{Y0}, \ref{bc}) defines set of 
parameters
\be
z_0\,=\,\infty\,; \quad C_1\,=\,0\,
; \quad C_2\,=\,0\,.\label{z0C}
\ee
However with these parameters the first integral in (\ref{eqF0}) diverges 
and we have no consistent solution. In view of this we consider the next 
approximation. We substioite solution (\ref{solution}) into~(\ref{eqF}) and 
calculate the term proportional to $\sqrt{z}$. Now we have

\beq
& &F(z)\,=\,1\,+\,\frac{g \sqrt{z}}{4\,\sqrt{3}\,\pi}\biggl(\ln(z)\,-\,4\,
\gamma\,-11\,\ln(2)\,+\,\frac{7}{12}\biggr)\,+\nn\\
& &+\,\frac{2}{3\,z}\,\int_0^z\,F(t)\,t\, dt\,
-\,\frac{4}{3\,\sqrt{z}}\,\int_0^z\,F(t)\,
\sqrt{t} dt\,-\nn\\
& &-\,\frac{4\,\sqrt{z}}{3}\,\int_z^{z_0}\,F(t)\,\frac{dt}{\sqrt{t}}\,+\,\frac
{2\,z}{3}\,\int_z^{z_0}\,F(t)\,\frac{dt}{t}\;
\label{eqFg}
\eeq
We look for solution of (\ref{eqFg}) in the form
\beq
& &F(z)\,=\,\frac{1}{2}\,G_{15}^{31}\Bigl( z\,|^0_{1,\,1/2,\,0,\,-1/2,\,-1}
\Bigr) -\,\frac{g \sqrt{3}}{16\,\pi}\,G_{15}^{31}\Bigl( z\,|^{1/2}_{1,\,1/2,
\,1/2,\,-1/2,\,-1}\Bigr)\,+\nn\\
& &+\,C_1\,G_{04}^{10}\Bigl( z\,|1/2,\,1,\,-1/2,\,-1\Bigr) + 
C_2\,G_{04}^{10}\Bigl( z\,|1,\,1/2,\,-1/2,\,-1\Bigr)\,; 
\label{solutiong}
\eeq
Boundary conditions also change to become in explicite form
\beq
& &\frac{1}{2}\,G_{04}^{30}\Bigl( z\,|1/2,\,0,\,-1/2,\,0\Bigr) -\,\frac{g 
\sqrt{3}}{16\,\pi}\,\biggl(\frac{1}{2}\,G_{04}^{30}\Bigl( z\,|1/2,\,0,\,0,
\,-3/2\Bigr)\,+\nn\\
& &+\,G_{15}^{31}\Bigl( z\,|^{-1}_{1/2,\,0,\,
0,\,0,\,-3/2}\Bigr)\biggr)+\,C_1\,G_{04}^{10}\Bigl( z\,|0,\,1/2,\,0,\,-1/2,
\Bigr) + \nn\\
& &+\,C_2\,G_{04}^{10}\Bigl( z\,|1/2,\,0,\,0,\,-1/2,\Bigr)\,-\nn\\
& &\,-\,\frac{1}{2 \sqrt{z}}\,-\,\frac{g \sqrt{3}}{4 \pi}\biggl(\ln(z) + \frac{9}{4} - 4 \gamma - 11\, \ln(2)\biggr)\,=\,0\,; \qquad z=z_0\,;\label{bcg}\\
& & \frac{1}{2}\,G_{04}^{30}\Bigl( z\,|1/2,\,0,\,-1/2,\,1\Bigr) 
-\,\frac{g 
\sqrt{3}}{16\,\pi}\,\biggl(\frac{1}{2}\,G_{15}^{31}\Bigl( z\,|^0_{1/2,\,0,\,
0,\,1,\,-3/2}\Bigr)\,+\nn\\
& &+\,G_{15}^{31}\Bigl( z\,|^{-2}_{1/2,\,0,\,
0,\,0,\,-3/2}\Bigr)\,-\,2\,G_{15}^{31}\Bigl( z\,|^{-1}_{1/2,\,0,\,
0,\,0,\,-3/2}\Bigr)\biggr)\,+\nn\\
& &+\,C_1\,G_{04}^{10}\Bigl( z\,|0,\,1,\,1/2,\,
-1/2,\Bigr) + 
C_2\,G_{04}^{10}\Bigl( z\,|1/2,\,1,\,0,\,-1/2,\Bigr)\,+\nn\\
& &+\,\frac{1}{4\, \sqrt{z}}\,-\,\frac{g \sqrt{3}}{4 \pi}\,=\,0\,; 
\qquad z=z_0\,;\nn
\eeq
Condition $F(0)=1$ leads to the following relation 
\be
1\,+\,\frac{G^2\,N}{64\,\pi^2}\,\int_0^{Y}\,F_0(y)\,y\,dy\,=\, 
\frac{3\,G\,g\,N}{16\,\pi^2}\,\int_0^Y\,F_0(y)\, dy\,;\label{g}
\ee
which in explicite form reads
\beq
& &1\,+\,8\,\biggl(\frac{1}{2}\,G_{26}^{32}\Bigl( z\,|^{1,\,1}_{2,\,3/2,\,1,
\,1/2,\,0,\,0}\Bigr) -\,\frac{g 
\sqrt{3}}{16\,\pi}\,\,G_{26}^{32}\Bigl( z\,|^{1,\,1/2}_{2,\,3/2,\,
3/2,\,1/2,0,\,0}\Bigr)+ \nn\\
& &+\,C_1\,G_{15}^{11}\Bigl( z\,|^1_{3/2,\,2,\,1/2,\,0,\,
0}\Bigr) +\,C_2\,G_{15}^{11}\Bigl( z\,|^1_{2,\,3/2,\,1/2,\,0,\,0}\Bigr)
\biggr)\,-\label{norm}\\
& & -\frac{3 \sqrt{3}\,g}{\pi}\,\Biggl(\frac{1}{2}\,G_{26}^{32}\Bigl( z\,|^
{1,\,1/2}_{3/2,\,1,\,1/2,
\,0,\,0,\,-1/2}\Bigr) -\,\frac{g 
\sqrt{3}}{16\,\pi}\,\,G_{26}^{32}\Bigl( z\,|^{1,\,1}_{3/2,\,1,\,
1,\,0,\,0,\,-1/2}\Bigr)+ \nn\\
& &+\,C_1\,G_{15}^{11}\Bigl( z\,|^1_{1,\,3/2,\,0,\,0,\,-1/2}\Bigr) +\,C_2\,
G_{04}^{10}\Bigl( z\,|3/2,\,3/2,\,0,\,0,\,-1/2\Bigr)\Biggr)\,=\,0\,;
\quad z\,=\,z_0\,.
\eeq
We have also condition
\be
F(z_0)\,=\,0\,;\label{pht1}
\ee
that means smooth transition from the non-trivial solution to trivial one 
$G\,=\,0$.
The conditions define values of parameters
\be
g(z_0)\,=\,2.55779\,;\quad z_0\,=\,1.915838\,;\quad 
C_1\,=\,0.06172743\,; \quad C_2\,=\,-0.1640803\,.\label{gY}
\ee
We would draw attention to the fixed value of parameter $z_0$. The solution 
exists only for this value~(\ref{z0C}) and it plays the role of eigenvalue.
As a matter of fact from the beginning the existence of such eigenvalue is by no 
means evident.

We consider the neglected terms of equation~(\ref{eqF}) as a perturbation to be 
taken into account in forthcoming studies. 

We use Schwinger-Dyson equation for gluon polarization operator to 
obtain a contribution of additional effective vertex to the running QCD 
coupling constant $\alpha_s$. The corresponding diagram is presented at 
Fig.2. Due to this vertex being gauge invariant, 
there is no contribution of ghost fields. So the contribution under 
discussion reads
\be
\Delta \Pi_{\mu \nu}(x)\,=\,\frac{g\,G\,N}{2\,(2\,\pi)^4}\int \frac{\Gamma^0_
{\mu \rho \sigma}(p,-q-\frac{p}{2},q-\frac{p}{2})\,\Gamma^{eff}_{\nu \rho \sigma}(-p,q+\frac{p}{2},-q+\frac{p}{2})\,F(q^2+\frac{3 p^2}{4})
\,dq}
{(q^2+p^2/4)^2\,-\,(p q)^2}\,.\label{SD}
\ee
Using expression (\ref{vertex}) after angular integrations we have
\beq
& &\Delta \Pi_{\mu \nu}(x)\,=\,(g_{\mu \nu}\,p^2 - 
p_\mu p_\nu)\,\Pi(x)\,;\quad x\,=\,p^2\,;\quad y'\,=\,q^2+\frac{3 x}{4}\,;\nn\\
& &\Pi(x)\,=\,-\, \frac{g\,G\,N}{32\,\pi^2}\Biggl(\frac{1}{x^2}\int_{3 x/4}^x  
\frac{F(y') dy'}{y'-x/2}\,\biggl(16\frac{y'^3}{x^2}-48\frac{y'^2}{x}+45 y-\frac
{27}{2}x\biggr)\,+\label{DF}\\
& &+\,\int_x^Y \frac{F(y') dy'}{y'-x/2}\,\biggl(-\,3 y'\,+\,\frac{5}{2}\,x
\biggr)\Biggr)\,.\nn
\eeq
Here coupling $g$ corresponds to $g(Y)$. Integrals of Meijer functions 
depending on $y'$ multiplied by any power of $y'$ can be evaluated 
analytically. In view of this we expand the denominator in~(\ref{DF}) in 
series in powers of $x/2$.  
Substituting solution~(\ref{solution}) into integrals we obtain explicit 
expressions for each term of expansion of expression~(\ref{DF}). 
Calculations show that results with different numbers of expansion terms 
 begin practically coincide starting from three-term expansion. In what 
follows  we present results with four terms.
 
So we have modified one-loop expression for $\alpha_s(p^2)$
\be
\alpha_s(x)\,=\,\frac{4\,\pi\,\alpha_s(p_0^2)}{4\,\pi+b_0\,\alpha_s(p_0^2)\,
\ln(x/\Lambda^2)\,+\,4\,\pi\Pi(x)}\,; \quad x=p^2\,; \quad b_0\,=\,11-
\frac{2\,N_f}{3}\,.\label{al1}
\ee
It is remarkable that function $\Pi(x)$ at $x\,=\,Y$ $(z = z_0)$ turns to be 
very small, 
almost zero. We just expect this quantity to be zero exactly. So this 
property of the approximated polarization operator indicates the consistency 
of the procedure being used. Namely 
we normalize  $\alpha_s(p^2)$ at point $p_0$, which correspond to 
our cut-off $Y$. Coupling constant $g$ entering in expressions~(\ref{g}) 
and~(\ref{DF}) is just corresponding to this normalization point. 
Performing the well-known transformations in expression~(\ref{al1}) 
we have for $u < u_0$ 
\beq
& & \alpha_s(u)\,=\,\frac{4\,\pi}{b_0}\Biggl(\ln(u)\,+\,\frac{8\,\sqrt{N}\,
\pi}{b_0\,g}\,I\Biggr)^{-1}\,; \quad u\,=\,\frac{x}
{\Lambda^2_{QCD}}\,;\quad u_0\,=\,\frac{Y}
{\Lambda^2_{QCD}}\,=\,14.61328\,;\nn\\
& & I\,=\,-\,\int_\frac{z}{16}^z \frac{
F_0(t) dt}{\sqrt{t}(2 \sqrt{t}-\sqrt{z})}\Biggl[16\frac{t^{3/2}}{z}-48\frac{t}{\sqrt{z}}+45 \sqrt{t}-\frac{27}{2}\sqrt{z}\Biggr]\,+\label{alpha1}\\
& &+\,\int_z^{z_0}\frac{F_0(t) dt}{\sqrt{t}(2 \sqrt{t}-\sqrt{z})}\Biggl[-3 
\sqrt{t}+\frac{5}{2}\sqrt{z}\Biggr]\,; \quad z=\frac{G^2\,N\,x^2}{1024\,\pi^2}\,;\quad b_0\,=\,9\,.\nn
\eeq
For $u > u_0$ we use the perturbative one-loop expression
\be
\alpha_s(u)\,=\,\frac{4\,\pi}{b_0\,\ln(u)}
\,.\label{pert1}
\ee
Two expressions~(\ref{pert1}),~(\ref{alpha1}) give equal values $\alpha_0$ for 
$\alpha_s$ at point $z_0$, while value 
$\alpha_s(u_0)$~(\ref{gY}) is defined by equation~(\ref{g}). 
The self-consistent result for expressions~(\ref{alpha1}, \ref{pert1}) with 
account of previous relations is unique and reads as follows
\beq
& &z_0\,=\,1.915838\,;\quad u_0\,=\,14.61328\,; \quad \alpha_0\,=\,0.5206188\,; 
 \nn\\
& & C_1\,=\,0.06172743\,; \quad C_2\,=\,-\,0.1640803\,.
\label{sol11}
\eeq
Behavior of $\alpha_s$~(\ref{alpha1}) with $u\,=\,Q/\Lambda_{QCD}$ is 
presented at Fig. 3 for  $\Lambda_{QCD}\,=\,0.2\,GeV$ and $0.05\,GeV\,<\,
Q\,<\,1\,GeV$. 
The behaviour with maximum at $Q \simeq 0.6\,GeV$ and maximal 
value $\alpha_s^{max} \simeq 0.55$ agrees to  
 calculations in work~\cite{Baldicchi}. Qualitately the result also 
corresponds to lattice calculations in work~\cite{Mueller} (see also 
discussion in 
paper~\cite{Shirk}). Note, that we begin plot at Fig.3 starting from 
$p\,=\,0.05\,GeV$, because $\alpha_s(u)$ has a pole at very small u, which 
is 
analogous to the well-known perturbative pole at $u\,=\,1$. Now this pole is 
shifted to the far infrared region. One may deal with it using the method 
proposed in work~\cite{Shirk97} and subtract from~(\ref{alpha1}) the 
following term
\be
\frac{4\,\pi}{b_0\, D\,(u\,-\,u_{00})}\,;\quad u_{00}\,=\,0.005769\,;\quad 
D\,=\,170.1594\,.\label{ShS}
\ee
This procedure practically does not change 
the result presented at Fig.3 in the denoted interval of $Q$. For comparison 
we present the modified $\alpha_s(Q)$ for interval $0.01\,GeV < Q < 1\,GeV$ 
 at Fig 4. Value of $\alpha_s(Q)$ at zero reads $\alpha_s(0)= 1.4205$.

All values are now expressed in terms of $\Lambda_{QCD}$. Emphasize, that 
there is no additional parameters to describe the non-perturbative infrared 
region.  

From definition of 
variables $z$, $u$ and $x$ we obtain value of new coupling constant $G$
\be
G\,=\,\frac{5.497571}{\Lambda_{QCD}^2}\,.\label{G1}
\ee

We calculate non-perturbative vacuum average of the third power in gluon 
field, which is immediately connected with our results. We have
\beq
& &<g^3\,f_{a b c}\,F^a_{\mu\nu}\,F^b_{\nu\rho}\,F^c_{\rho\mu}>\,=\,
\frac{g^3\,G\,96\,\pi^4}{(2\pi)^8}\,I_1\,I_2\,=\,110.77\,\Lambda^6_{QCD}\,;
\label{F3}\\
& &I_1\,=\,\int_0^\frac{4}{3}\biggl(1-\frac{3 y}{4}\biggr) 
dy-\int_0^{1}\frac{(1- y)^2}{(1-y/2)} dy-\int_1^\frac{4}{3}\frac{(1-y)^2(1-3 y/4)}{y(1-y/2)} dy\,=\nn\\
& &=\,0.278756\,;\quad I_2\,=\,\int_0^Y x^3\,dx\,F(x)\,=\,0.18062\,\frac{(32\,
\pi)^4}{2\,G^4\,N^2}\,.\nn
\eeq
Value~(\ref{F3}) seems reasonable. However the accuracy of phenomenological 
and lattice definitions of this value is not yet satisfactory.

To conclude we would state, that method~\cite{Arb04} being applied to 
non-perturbative $\alpha_s$ proves its efficiency even in the first 
approximation, which is considered here. Bearing in mind also results of 
works~\cite{Arb05, AVZ} on application of the approach to low-energy 
hadron physics we would express a hope, that in this way we 
could obtain the adequate tool to deal non-perturbative effects in 
QCD and, maybe, in other problems.

The mean non-perturbative value for $\alpha_s$ turns to be here 
around $0.5$. For this value results of works~\cite{Arb05, AVZ} seem to 
be quite consistent. 

The author express his gratitude to D.V. Shirkov for valuable discussions.

\newpage
\begin{center}
{\bf Figure captions}
\end{center}
\bigskip
\bigskip
Fig.1. Diagram representation of the compensation 
equation. Black spot corresponds to tree-gluon  
vertex with a form-factor. Simple point corresponds to 
a point-like vertex. Incoming momenta are denoted by the corresponding 
external lines.\\
\\
Fig.2. Loop contribution to polarization operator.\\
\\ 
Fig.3. Behaviour of $\alpha_s(Q)$, for $0.05\,GeV < Q < 1\,GeV$, 
$\Lambda_{QCD}\,=\,0.2\,GeV$. 
\\
Fig.4. Behaviour of modified $\alpha_s(Q)$, for $0.01\,GeV < Q < 1\,GeV$, 
$\Lambda_{QCD}\,=\,0.2\,GeV$. 

\newpage
\begin{picture}(160,75)
{\thicklines
\put(5,60.5){\line(-3,2){10}}
\put(5,60.5){\line(-3,-2){10}}
\put(5,60.5){\circle*{3}}}
\put(5,60.5){\line(1,0){13}}
\put(-5,70.5){p}
\put(-5,50.5){-p}
\put(10,62.5){0}
\put(23.5,60){+}
{\thicklines
\put(52.5,60.5){\line(-3,2){15}} 
\put(52.5,60.5){\line(-3,-2){15}}
\put(37.5,72.5){p}
\put(37.5,47.5){-p}
\put(57.5,62.5){0}
\put(52.5,60.5){\circle*{3}}
\put(42.5,53.5){\line(0,1){13}}} 
\put(52.5,60.5){\line(1,0){13}}}
\put(80,60){+}
{\thicklines
\put(102.5,60.5){\line(-3,2){10}}
\put(102.5,60.5){\line(-3,-2){10}}
\put(112.5,60.5){\oval(20,10)}
\put(122.5,60.5){\line(1,0){13}}
\put(122.5,60.5){\circle*{3}}
\put(92,70.5){p}
\put(92,50.5){-p}
\put(127.5,62.5){0}
\put(80,60){+}
{\thicklines
\put(102.5,60.5){\line(-3,2){10}}
\put(102.5,60.5){\line(-3,-2){10}}
\put(112.5,60.5){\oval(20,10)}
\put(122.5,60.5){\line(1,0){13}}
\put(122.5,60.5){\circle*{3}}}
\put(0,20){+}
{\thicklines
\put(32.5,30.5){\line(-3,2){10}}
\put(32.5,10.5){\line(-3,-2){10}}
\put(32.5,20.5){\oval(10,20)}
\put(32.5,10.5){\line(1,0){13}}
\put(32.5,30.5){\circle*{3}}
\put(22.5,40.5){p}
\put(22.5,0){-p}
\put(40.5,12.5){0}}
\put(60,20){+}
{\thicklines
\put(92.5,10.5){\line(-3,-2){10}}
\put(92.5,30.5){\line(-3,2){10}}
\put(92.5,20.5){\oval(10,20)}
\put(92.5,30.5){\line(1,0){13}}
\put(92.5,10.5){\circle*{3}}
\put(82.5,40.5){p}
\put(82.5,0){-p}
\put(100.5,32.5){0}}
\put(120,20){=}
\put(130,20){0}
\end{picture}

\bigskip
\bigskip
\bigskip
\bigskip
\bigskip

\begin{center}
Fig. 1. 
\end{center}
\newpage
\begin{picture}(160,40)

{\thicklines
\put(62.5,20.5){\line(-1,0){15}} 
\put(72.5,20.5){\oval(20,10)}
\put(82.5,20.5){\line(1,0){15}}
\put(55.5,22.5){p}
\put(82.5,20.5){\circle*{3}}
\put(90.5,22.5){-p}}
\end{picture}

\begin{center}
Fig. 2. 
\end{center}

\newpage

\begin{figure}[ht]
\begin{picture}(300,450)
\put(20,300){\epsfig{file=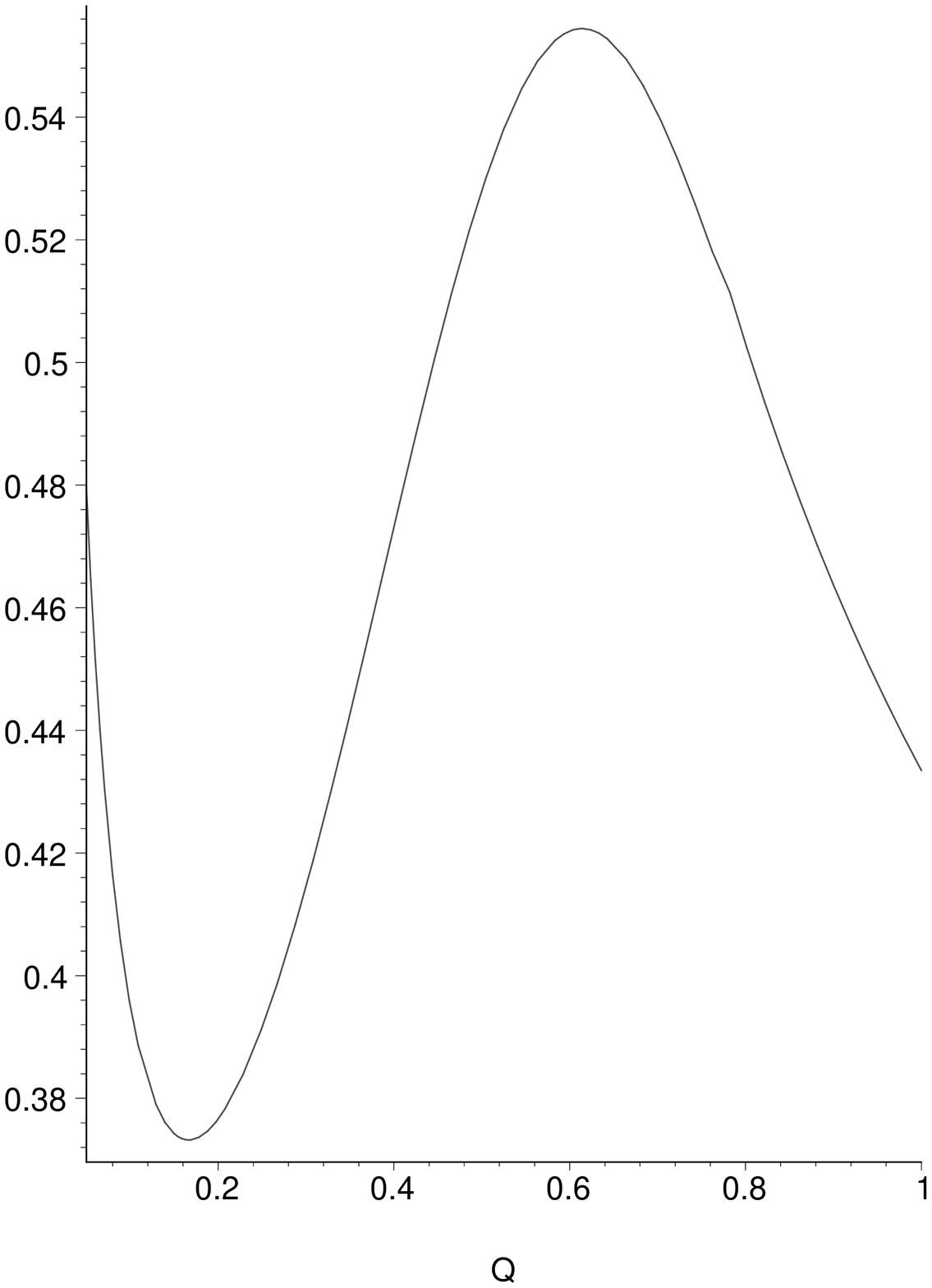,width=300pt, height=450pt}} 
\put(60,270){Fig. 3}
\put(5,455){$\alpha_s(Q)$}
\put(130,310){$GeV$}
\end{picture}
                                                                               
\caption{}
                                                                                
\label{}
                                                                                
\end{figure}

\newpage

\begin{figure}[ht]
\begin{picture}(300,450)
\put(20,300){\epsfig{file=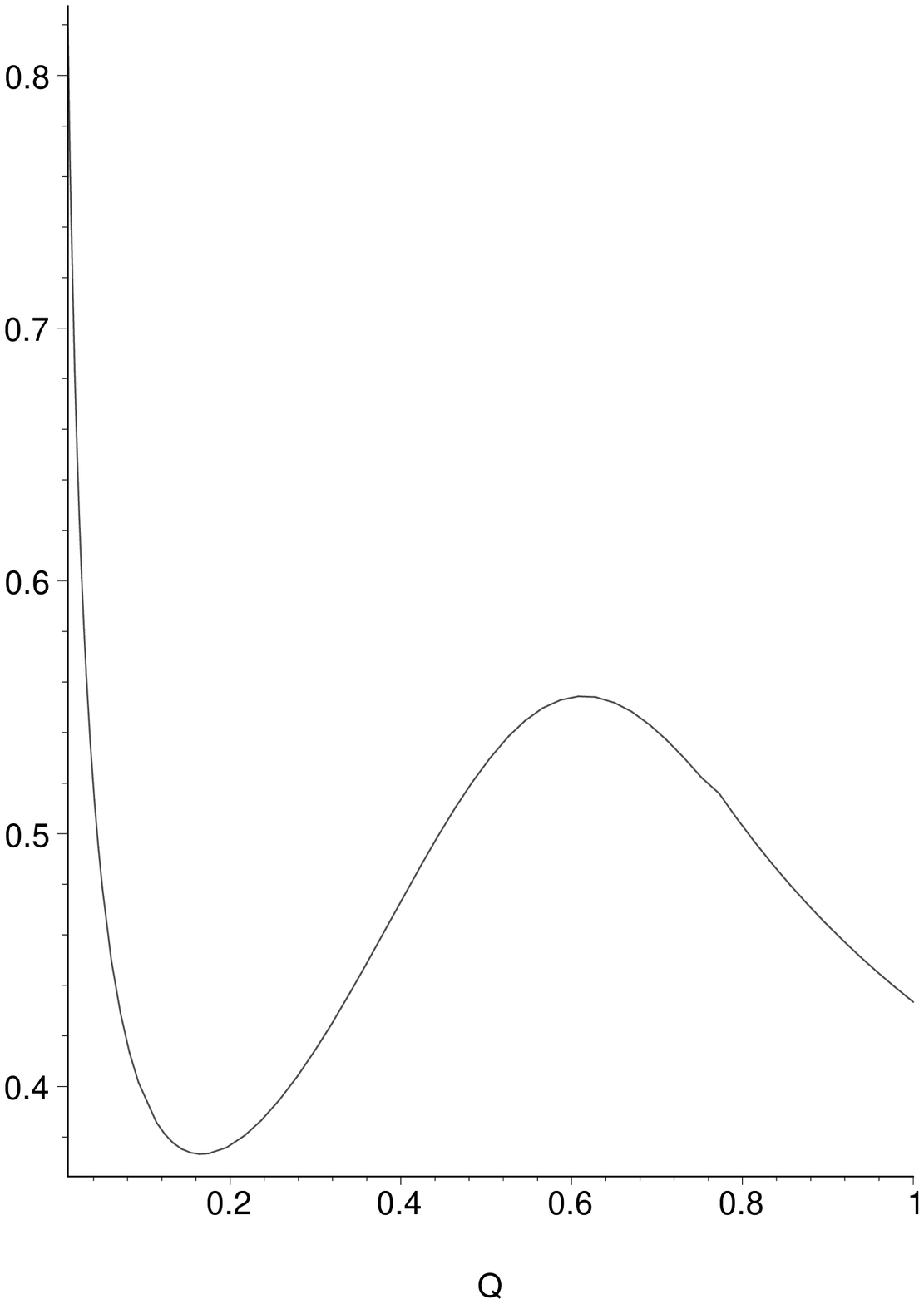,width=300pt, height=450pt}} 
\put(60,270){Fig. 4}
\put(5,455){$\alpha_s(Q)$}
\put(130,310){$GeV$}
\end{picture}
                                                                               
\caption{}
                                                                                
\label{}
                                                                                
\end{figure}

\end{document}